\documentclass[epj,twocolumn]{webofc}
\usepackage[varg]{txfonts}

\usepackage{subfigure}
\usepackage{nth}
\usepackage{soul}

\usepackage[english]{babel}
\usepackage[numaddn=\pi,per=fraction]{siunitx}
\usepackage{graphicx}

\newcommand{\imgsize}{0.5\textwidth}

\usepackage{amssymb}

\woctitle{International Symposium on Very High Energy Cosmic Ray Interactions 2014}

\begin{document}

\title{Measuring the Muon Content of Air Showers with IceTop}

\author{Javier G. Gonzalez \inst{1} for the IceCube Collaboration}

\institute{University of Delaware}

\abstract{
IceTop, the surface component of the IceCube detector, has been used
to measure the energy spectrum of cosmic ray primaries in the range
between \SI{1.58}{PeV} and \SI{1.26}{EeV}. It can also be used to
study the low energy muons in air showers by looking at large
distances (>~300m) from the shower axis. We will show the muon lateral
distribution function at large lateral distances as
measured with IceTop and discuss the implications of this
measurement. We will also discuss the prospects for low energy muon
studies with IceTop.
}

\maketitle


It is well known that the muon content of an air shower, together with
a measure of its electromagnetic component, can be used to estimate
the energy and mass of its primary \cite{Kampert:2012mx}. The main
issue with the use of the muon content as an estimate of primary mass
is the possible systematic differences between simulated and real air
showers, arising from the lack of knowledge of high energy hadronic
interactions. An excess number of muons has been reported by the HiRes-MIA and the Pierre Auger collaborations
\cite{PhysRevLett.84.4276,Aab:2014pza}. Understanding the systematic differences in air
shower muon content between simulations and data is one of the
pressing issues in the physics of very high energy cosmic rays.

The IceTop detector is sensitive to the low-energy (E $\gtrsim$ 200 MeV) muon component of
air showers. Generally speaking, at low zenith
angle and close enough to the air shower axis, the signal from muons
is overwhelmed by the signal from the electromagnetic (EM) component
of the air shower (electrons, positrons and gamma-rays). This holds
true in the zenith angle and lateral distance ranges that have been
used in the cosmic ray spectrum determination with IceTop ( $\theta <$
\SI{40}{\degree}, r $\lesssim 220 \cdot (E/PeV)^{1/4} $ m).  We are
interested in determining the average muon Lateral Distribution
Function (LDF) at lateral distances where the signal from muons
becomes significant. The lateral distance of any point is defined as
the closest distance from the point to the shower axis.

The estimation of the muon lateral distribution hinges on being able
to distinguish the signal produced by single muons from the signal
produced by EM particles. IceTop detector stations accomplish this by
virtue of their size. A typical electron leaves a signal proportional
to its track length through the detector, which is on the order of 10
cm. A typical muon crosses the detector, producing a track typically
longer than 90 cm. After describing the general features of IceTop
signals in Section \ref{section:general}, we will describe a method
for measuring the number of muons in Section \ref{section:ldf}. This
method does not rely on counting muons, but it is a fit to the signal
distributions within small bins of lateral distance, primary energy and
zenith angle.

\section{General Features of IceTop}
\label{section:general}

The IceCube detector consists of two major components.  It can
measure air showers on the surface with IceTop, high energy muon
bundles with the in-ice detector, and both components in coincidence
provided that the air shower axis goes through the in-ice detector. In what
follows, we will consider the specific characteristics of IceTop that
are relevant for measuring the low-energy muon component of air
showers. A more detailed description of IceCube and IceTop has already been presented elsewhere~\cite{IceCube:2012nn}.

IceTop is an air shower array consisting of 81 stations forming a
triangular grid with a separation of \SI{125}{m} in its completed
configuration. The results presented here were obtained with data
collected between June 1st 2010 and May 13th 2011, when IceTop
consisted of 73 stations. It is located above the deep IceCube
detector at the geographical South Pole, covering an area of roughly one
square kilometer. Each station consists of two ice Cherenkov tanks
separated by ten meters.  Each tank contains two Digital Optical
Modules (DOMs) with a 10 inch photomultiplier tube (PMT) and
electronics for signal processing and readout. 
A discriminator trigger occurs when the voltage in one of the DOMs in
a tank has passed the discriminator threshold. The total charge
collected at the PMT's anode, after digitization and baseline subtraction,
constitutes the tank's signal. The tanks register
signals ranging from 0.2 to 1000 Vertical Equivalent Muons
(VEM). A Hard Local Coincidence (HLC) occurs when
there are discriminator triggers in two neighboring tanks
within a time window of \SI{1}{\micro s}. If there is a discriminator
trigger but not an HLC, the result is a Soft Local Coincidence (SLC).

\begin{figure}
  \includegraphics[width=\imgsize]{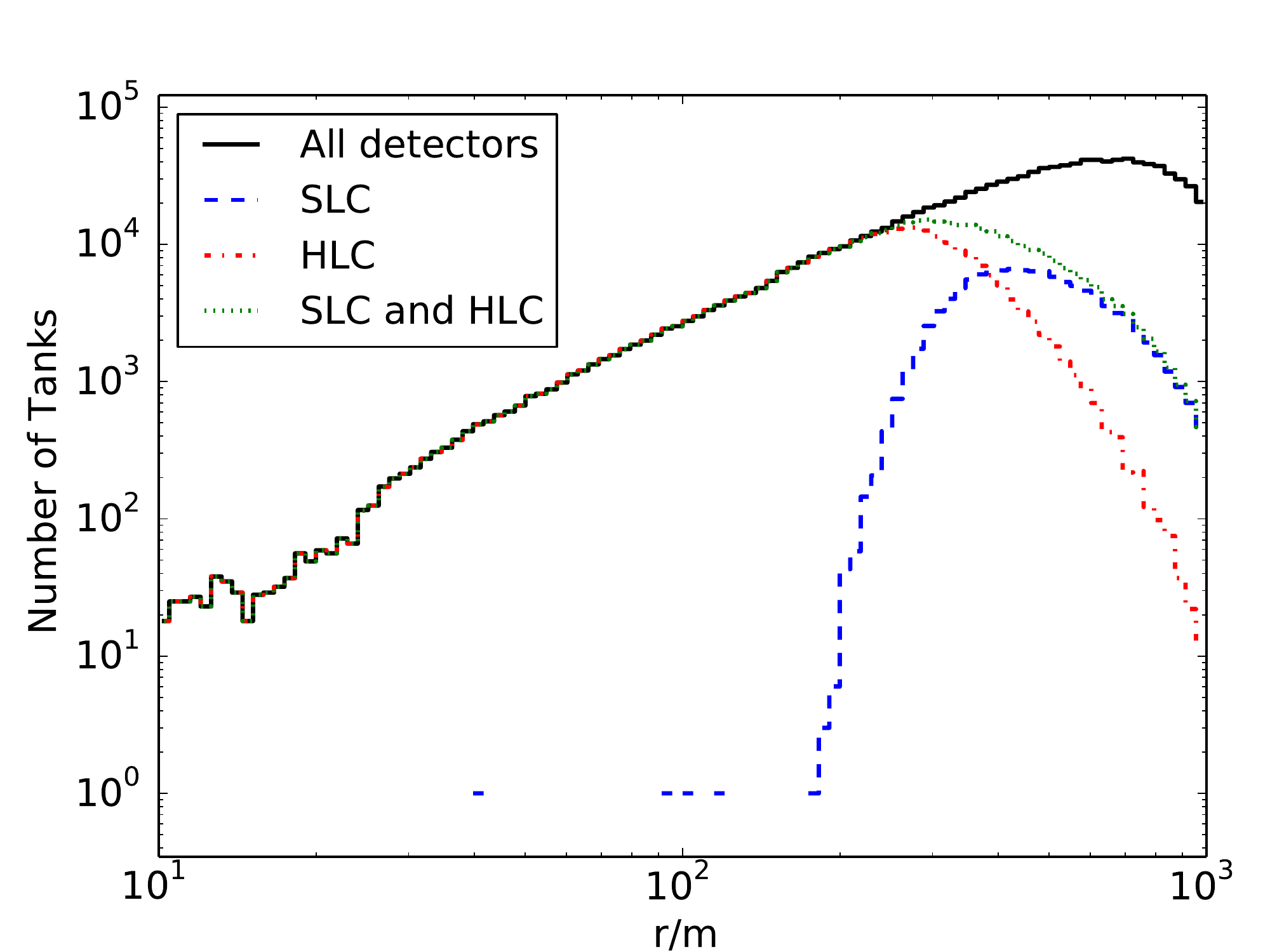}
  \caption{Lateral distribution of tanks in showers arriving with zenith angles less than
    \SI{6}{\degree} and with energies between \SI{10}{PeV} and \SI{12.6}{PeV}. The tanks
    are classified depending on whether they registered \textit{SLC}
    or \textit{HLC} signals (described in the text). The upper curve
    includes all tanks, including tanks that registered no
    signal during the event. In determining the muon content, we will only consider tanks at lateral
    distances larger than where the SLC and HLC lines cross. }
  \label{fig:tank_distributions}
\end{figure}

All analyses of IceTop data up to now have only considered signals where
both tanks in a station pass the threshold, or HLC signals. In this contribution we also consider SLC signals,
where the partner tank within the station did not have a discriminator trigger. SLC signals occur at
large lateral distances, where the triggering probability is
smallest. An example of the lateral distribution of SLC and HLC
signals from experimental data is shown in Figure~\ref{fig:tank_distributions}.

\begin{figure}
  \includegraphics[width=\imgsize,keepaspectratio=true]{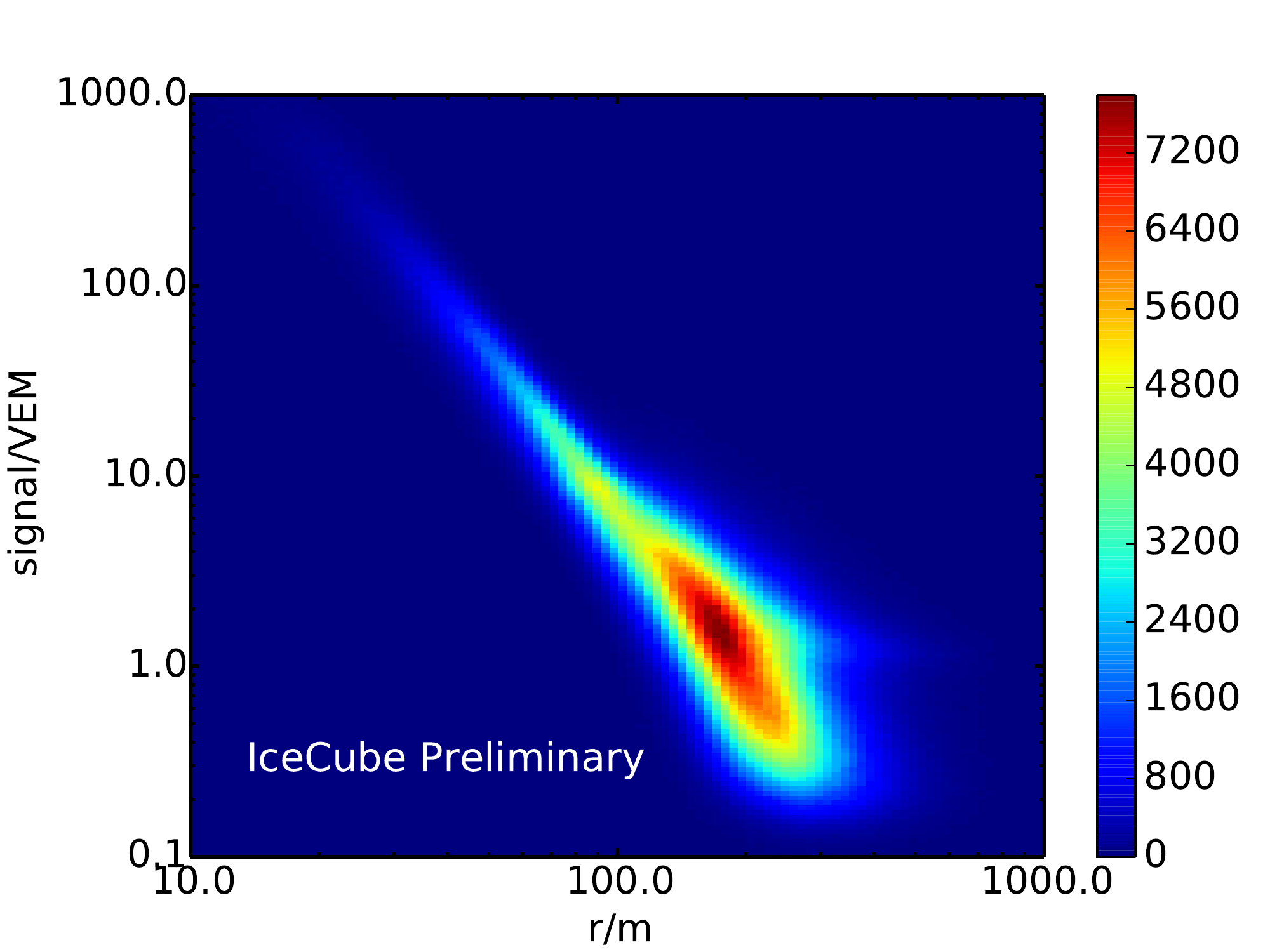}
  \caption{HLC signal distribution as a function of lateral distance
    for air showers with energies between 4 and 5 PeV, and zenith angle
    between \SI{28}{\degree} and \SI{32}{\degree}. This distribution
    is a 2d histogram that includes all HLC signals from all events
    with the given energy and arrival direction.}
  \label{fig:ldf_hlc}
\end{figure}

The
properties of the primary cosmic ray are reconstructed by fitting the
measured signals with a Lateral Distribution Function (LDF) which
includes an attenuation factor due to the snow cover on top of each
tank. For a given primary energy and arrival direction, the observed lateral distribution of signals used in this
reconstruction is shown in Figure~\ref{fig:ldf_hlc}. The signal
times are fitted with a function describing the shape of the shower
front.  The primary energy is given by the shower size S$_{125}$, defined as the
signal at a lateral distance of \SI{125}{m}.  The resulting
cosmic ray energy spectrum measured with IceCube/IceTop has been
presented previously \cite{Abbasi:2012wn,icetop_2013,icetop_coinc_2013}.
In the shower reconstruction just outlined, only HLC signals were considered.

The distinction between SLC and HLC signals provides a natural way to
identify tanks where we expect to see a significant muon
contribution. Generally speaking, we expect that signals at large
lateral distances will be mostly due to muons, whereas the signals at
short lateral distances will be mostly due to electrons and
$\gamma$-rays. Instead of relying on simulations for selecting signals
where the contribution from muons dominate, we will select tanks
where SLC signals are most likely. Specifically, we will restrict
ourselves to tanks beyond a lateral distance at which SLC signals
amount to 50\% of the total number of signals. The exact lateral
distance at which this happens depends on the energy of the primary. The
dependence on zenith angle is very weak. Therefore, we chose to use
the value corresponding to \SI{30}{\degree}.

\begin{figure}
  \includegraphics[width=\imgsize,keepaspectratio=true]{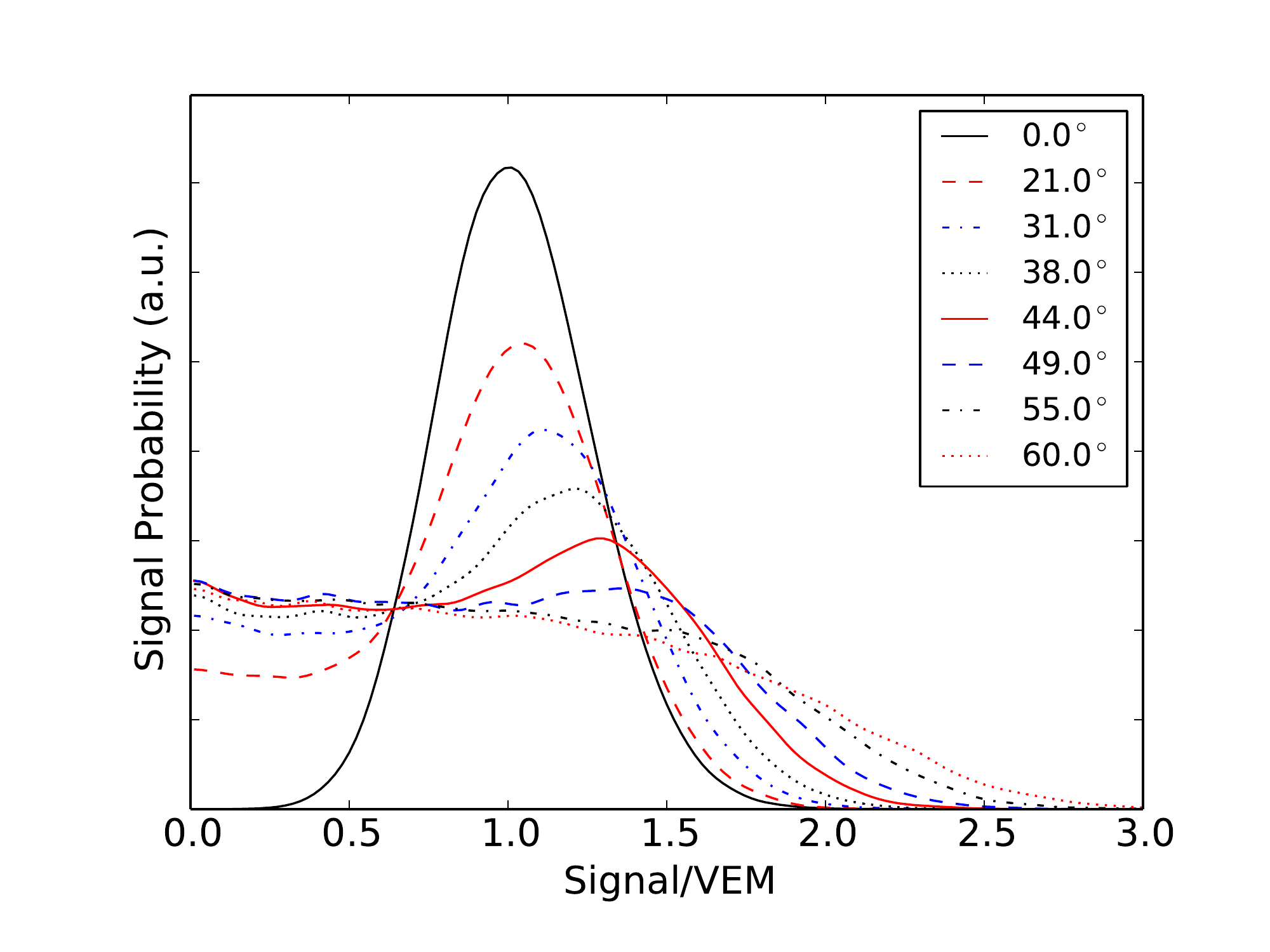}
  \caption{Signal distributions resulting from simulating the detector
    response for single muons arriving at fixed zenith angles from \SI{0}{\degree} to \SI{57}{\degree}.}
  \label{fig:single_muons}
\end{figure}

\begin{figure}
  \includegraphics[width=\imgsize]{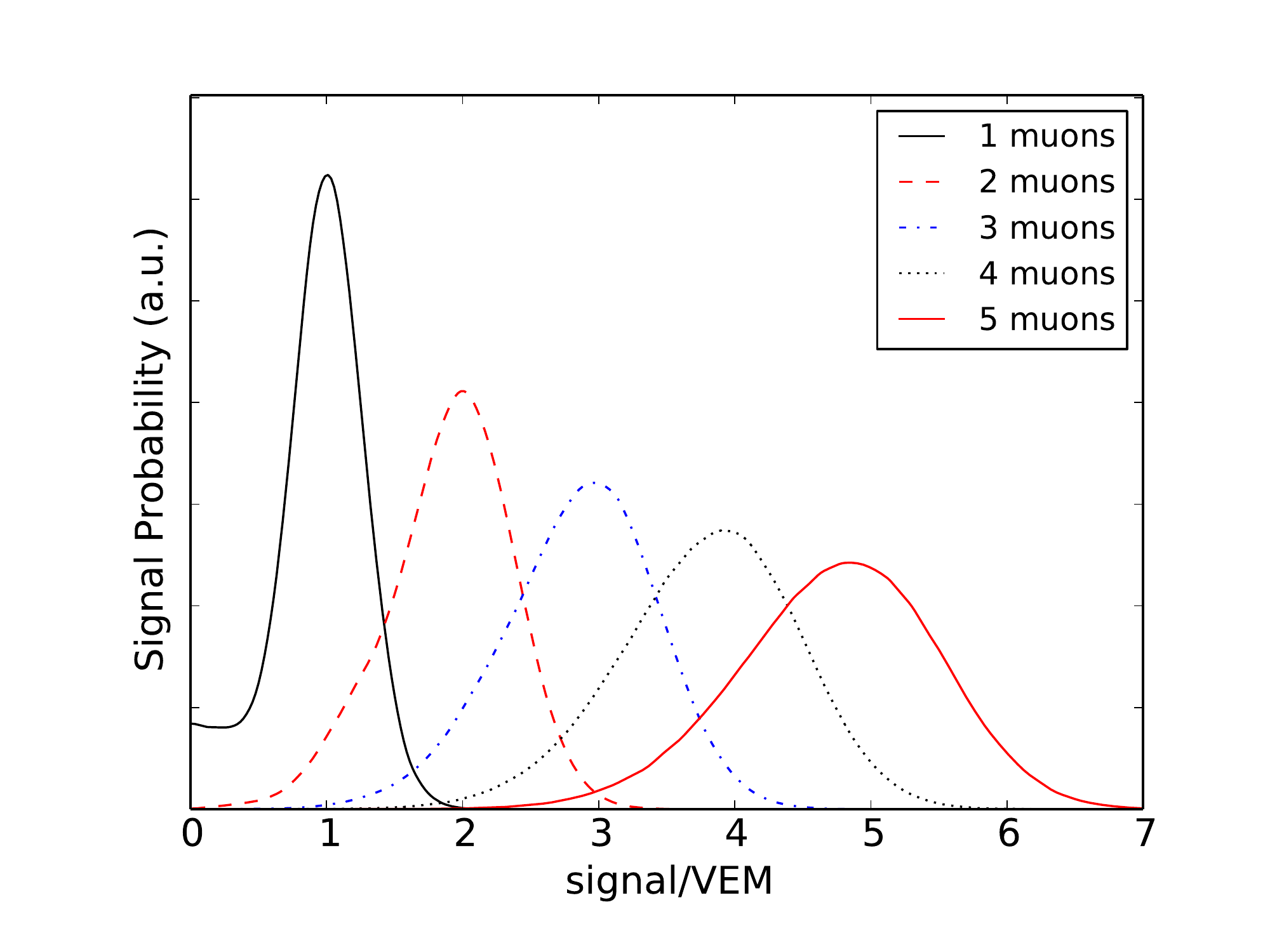}
  \caption{Signal distributions resulting from simulating the detector
    response for an integer number of muons arriving arriving at \SI{10}{\degree}.}
  \label{fig:many_muons}
\end{figure}

The statistical distribution of signals from EM particles will roughly
mimic their energy distribution, with a mean signal that corresponds
to a few tens of centimeters of track length inside the tank. On
the other hand, the distribution of signals from muons is mainly
determined by the geometry of the tank. The signal distributions
produced by single muons are obtained using the Geant4
toolkit~\cite{geant4}. Example distributions at various incident angles are
displayed in Figure~\ref{fig:single_muons}. The distribution is
clearly not symmetric. The peak of the distribution corresponds to
muons that enter through the top of the tank and exit through the
bottom. By definition, the peak position for vertically through-going
muons is one \textit{Vertical Equivalent Muon} (VEM). For muons
arriving at a zenith angle $\theta$, the peak is at
$1/\cos(\theta)$. The flat part at low signals corresponds to muons
with a short track through the tank, what we call \textit{corner
  clipping} muons. At large angles, few muons go through top and
bottom. For an integer number of muons, the signal distribution is just
the multiple auto-convolution of the single-particle distribution. An
example of this is displayed in Figure \ref{fig:many_muons}. The total
signal distribution for an expected average number of muons $\langle
N_{\mu} \rangle$ is given by a linear combination of the signal
distributions for integer numbers of muons:

\begin{equation}
\label{eq:muon_signal}
p\left(s \, \vert \, \langle N_{\mu} \rangle \right) = \sum_{n=0}^{\infty} \frac{\langle N_{\mu} \rangle^n}{n!} e^{-\langle N_{\mu} \rangle} p\left(s \,|\, N_{\mu}=n \right)
\end{equation}


\section{Determining the Muon Lateral Distribution}
\label{section:ldf}

\begin{figure}
   \includegraphics[width=\imgsize,keepaspectratio=true]{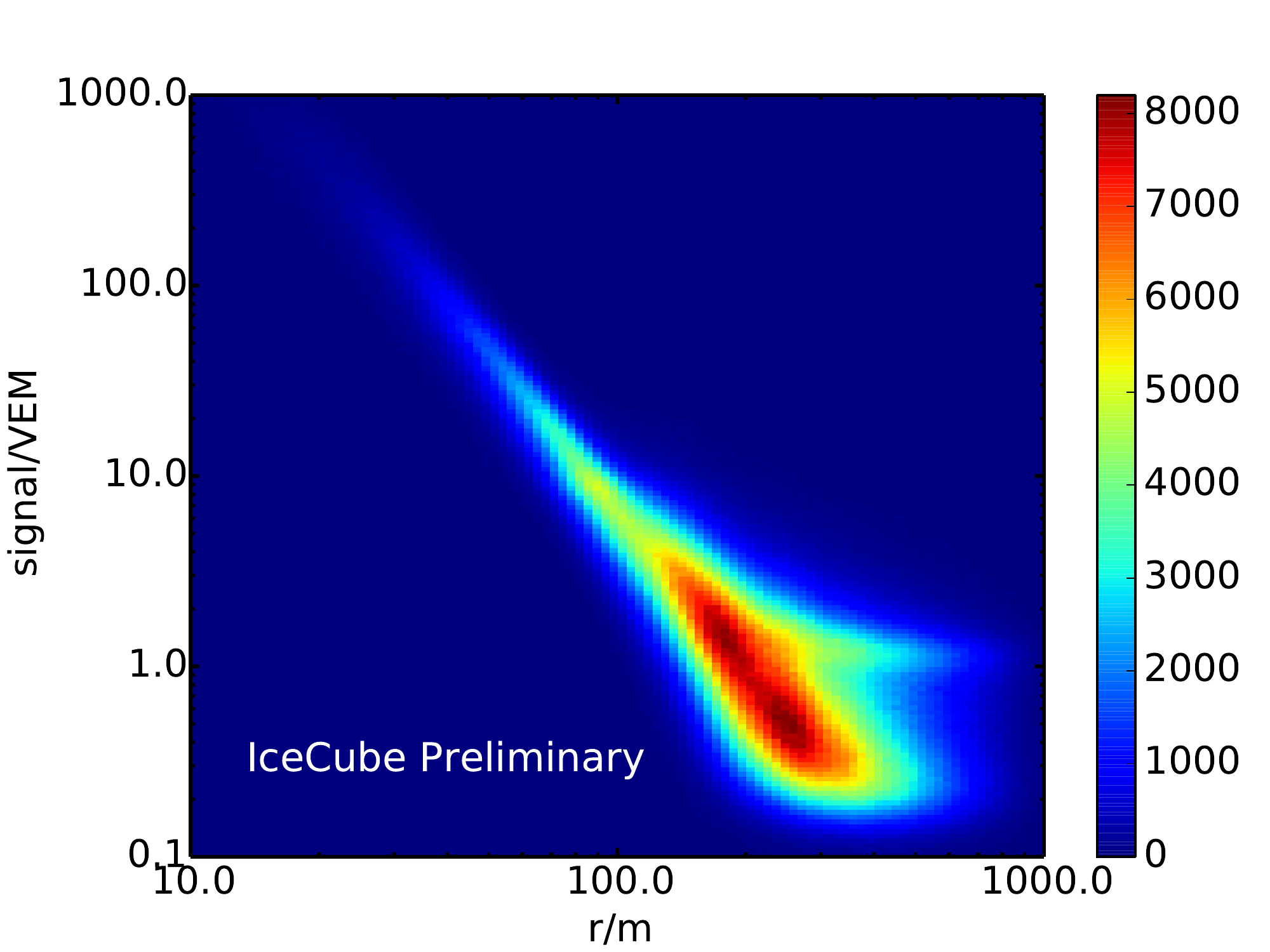}
   \caption{Signal distribution as a function of lateral distance for
     air showers with energies between 4 and 5 PeV, and zenith angle
     between \SI{28}{\degree} and \SI{32}{\degree}. This distribution
     includes all signals (SLC and HLC). The vertical lines mark the
     distances at which the 1-d histograms in
     Figure~\ref{fig:r_signal_slices} were made.}
   \label{fig:ldf}
 \end{figure}

\begin{figure}
   \includegraphics[width=\imgsize]{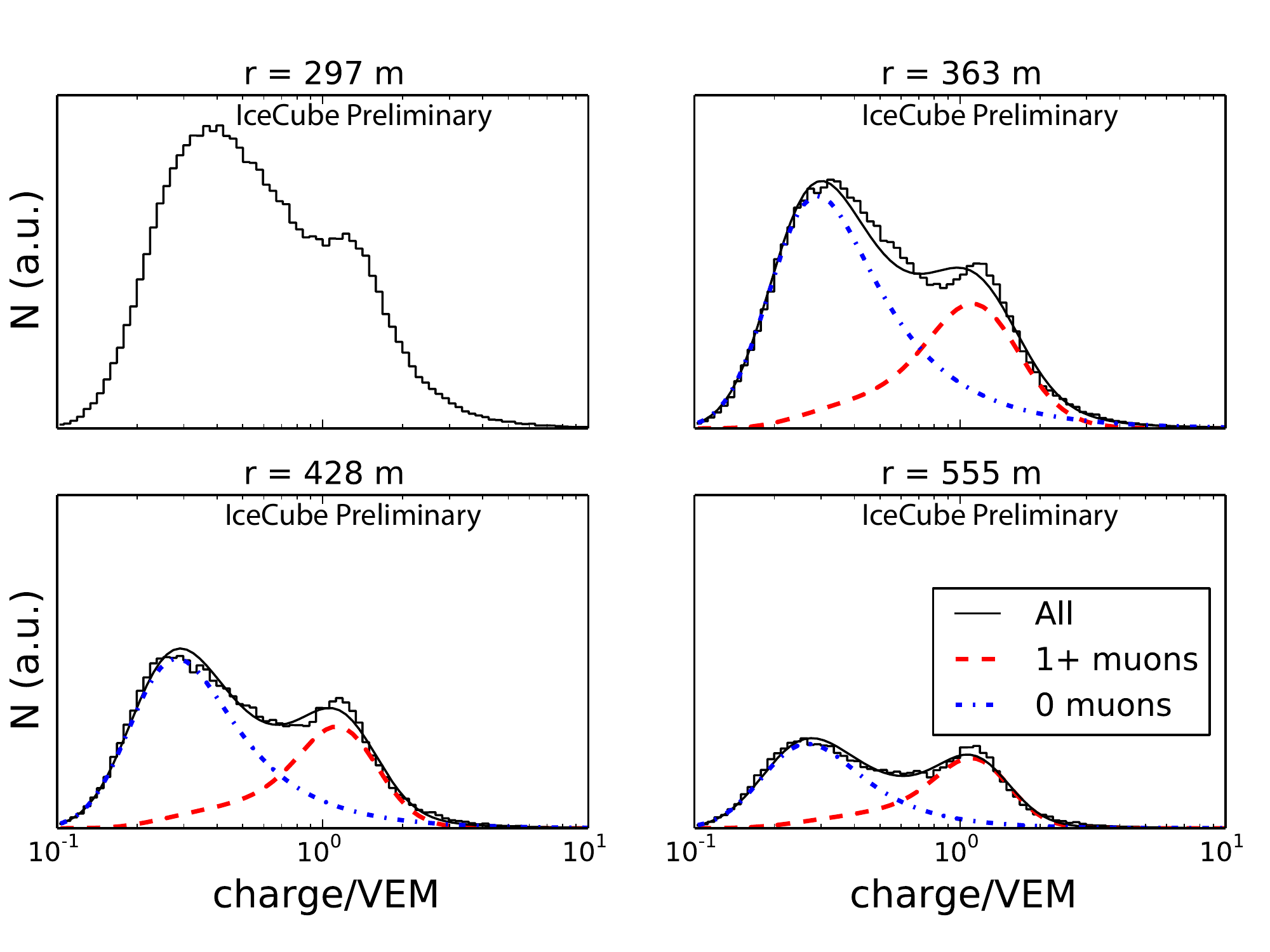}
   \caption{Vertical slices of the 2-d histogram in
     Figure~\ref{fig:ldf}. Each histogram corresponds to a vertical
     line in Figure~\ref{fig:ldf}. The fit lines are not shown at
     \SI{297}{m} because at this point the assumption made of the
     shape of the EM signal, that the peak EM signal is below
     threshold, is not valid. }
   \label{fig:r_signal_slices}
 \end{figure}

An example of the observed distribution of IceTop signals, corresponding
to air showers with fixed energy and zenith angle, can be seen in
Figure~\ref{fig:ldf}, where we now included the SLC signals. At large
distances, there are two distinct populations. One population is the
continuation of the main distribution at smaller distances, which
roughly follows a power law, where the electromagnetic component of
the shower dominates. The other population, with signals around 1 VEM,
is made up mostly of tanks hit by one or more muons. These two
populations are clearly seen in Figure~\ref{fig:r_signal_slices}, where
we show the histograms of signals registered at selected fixed lateral distances.

The first population corresponds to tanks hit by no muons. The entire
signal in these tanks is produced by electrons or
$\gamma$-rays. We approximate this distribution by a power-law
multiplied by a function that describes the trigger probability. The
trigger probability can be described by a sigmoidal function of the
logarithm of the charge, centered at 0.25 VEM and with a width of
0.14. Clearly, this approximation only works well at large lateral distances,
where the mean expected signal is well below
threshold and we thus look into the tail of the signal
distribution. The second population, with a peak around
\SI{1}{VEM}/$\cos\theta$, can be described by the contributions of
tanks hit by an integer number of muons. The charge distribution
obtained from simulations was described in Section
\ref{section:general} and is given by Equation \ref{eq:muon_signal}. This distribution is smeared and shifted to
account for a very small contribution from electrons, positrons, and
$\gamma$-rays.


We determine the muon lateral distribution function by independently
fitting the charge distributions at fixed energy, zenith, and lateral
distance, like the distributions shown in
Figure~\ref{fig:r_signal_slices}. The result of the fit is the number
of tanks hit by at least one muon. This, together with the total
number of tanks located at that distance (the upper curve in
Figure~\ref{fig:tank_distributions}), provides an estimate of the
probability that a tank is hit by one or more muons, which leads
to the mean number of muons $\langle N_{\mu} \rangle$:

\begin{equation}
p_{\mu\,hit} = \frac{N_{tanks\,with\,muons}}{N_{all\,tanks}} = 1 - e^{-\langle N_{\mu} \rangle}.
\end{equation}

The mean number of muons is divided by the cross-sectional area
of the tanks to yield the muon density at that location. In doing
this, we assume that the direction of motion of all the muons coincide
with the reconstructed air shower direction.

\begin{figure}
  \includegraphics[width=\imgsize]{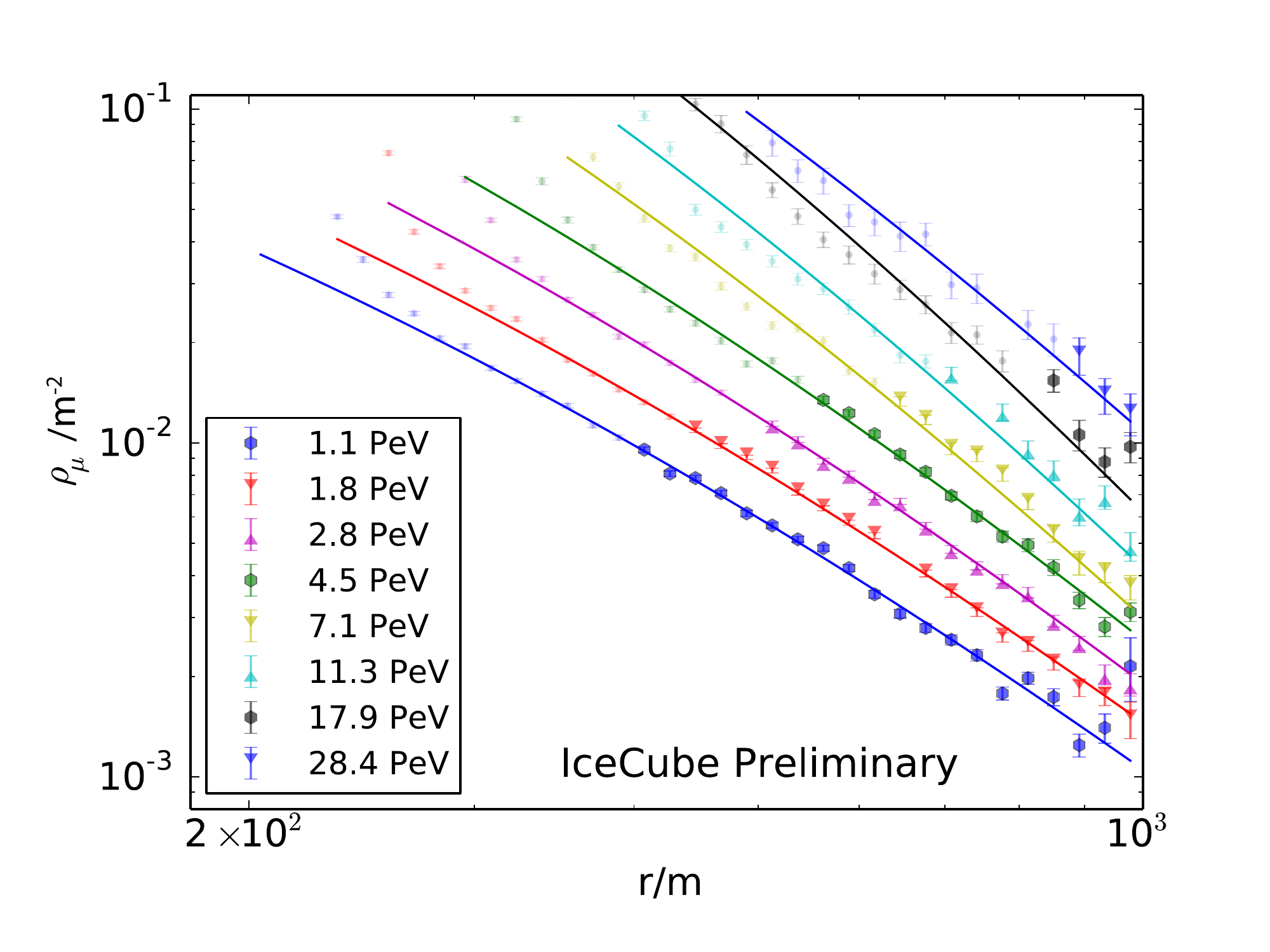}
  \caption{Reconstructed average lateral distribution of muons for air
    showers arriving at zenith angles of less than \SI{6}{\degree}
    and selected S$_{125}$ values.}
  \label{fig:muon_ldf_0}
\end{figure}

\begin{figure}
  \includegraphics[width=\imgsize]{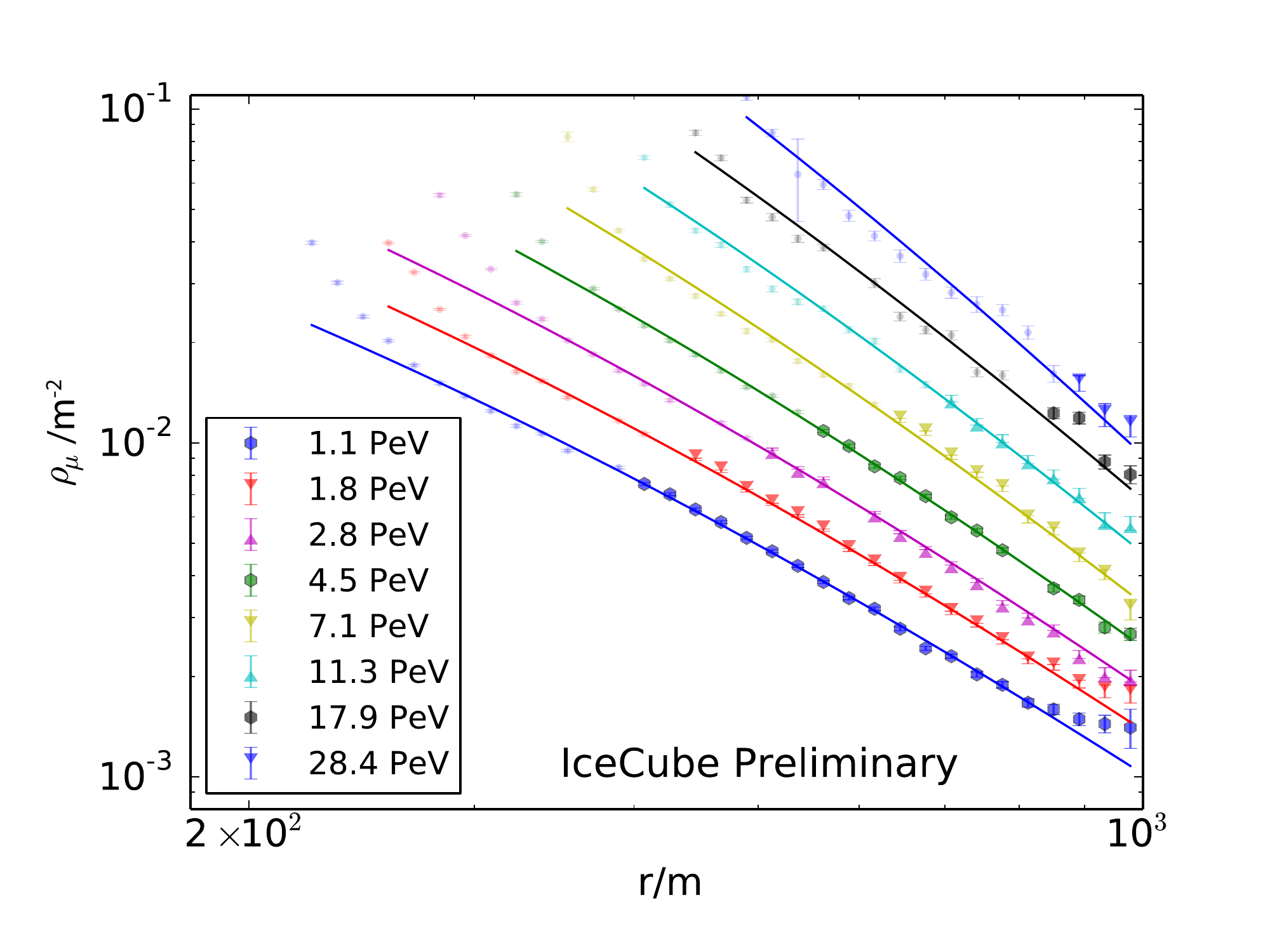}
  \caption{Reconstructed average lateral distribution of muons for air
    showers arriving at zenith angles between \SI{28}{\degree} and \SI{31}{\degree},
    and selected S$_{125}$ values.}
  \label{fig:muon_ldf_30}
\end{figure}

The resulting muon density for each lateral distance and some of the
energies considered, at two different zenith angles ($0^\circ$ and
$31^\circ$ from the vertical), are shown in
Figures~\ref{fig:muon_ldf_0} and \ref{fig:muon_ldf_30}. At this point
we should remember that all the points in these figures are chosen so
the number of SLC signals at any of the considered lateral distances
are at least 50\% of the number of signals at that lateral distance.
The smaller and fainter markers correspond to tanks located at
lateral distances in which SLC signals amount to less than 80\% of all
signals at that distance. Clearly, increasing this lateral distance
cut enhances the muon's contribution to the signal, but at the same
time limits the accuracy of the lateral distribution fit. The
optimization of this cut remains under study.

Each LDF can be described by the following function~\cite{Greisen_1960_review}:
\begin{equation}
\rho_\mu(r) = \rho_{\mu}(r_0) \, \left(\frac{r}{r_0}\right)^{-3/4} \left( \frac{320\,\text{m} + r\,}{320\,\text{m} +r_0}\right)^{-\gamma},
\end{equation}
which displays the same functional form as Greisen's
function, with the first exponent of r fixed to -3/4. The parameters
that are fitted are $\gamma$ and $\rho_{\mu}(r_0)$, which represents
the density of muons at a given lateral distance $r_0$. We have chosen
600 meters as the value for $r_0$. The choice of $r_0$ is arbitrary
and is motivated by the fact that the signals shown in
Figures~\ref{fig:muon_ldf_0} and \ref{fig:muon_ldf_30} are at lateral distances
between 300 and 1000 meters. It is also convenient because this is the
lateral distance at which previous experiments, notably Akeno and
HiRes-MIA \cite{Akeno_1984,Akeno_muons_1995,PhysRevLett.84.4276},
reported their results. It must be noted that the optimum value for
$r_0$ depends on energy and on the lateral distance cut. The two
parameters, $\rho_{\mu}(r_0)$ and $\gamma$, potentially depend on energy
and zenith angle. In the following section we will show how
$\rho_{\mu}(r_0)$ depends on energy in the case of vertical air showers.

\section{Results and Discussion}
\label{section:results}

\begin{figure}
  \begin{center}
    \includegraphics[width=\imgsize]{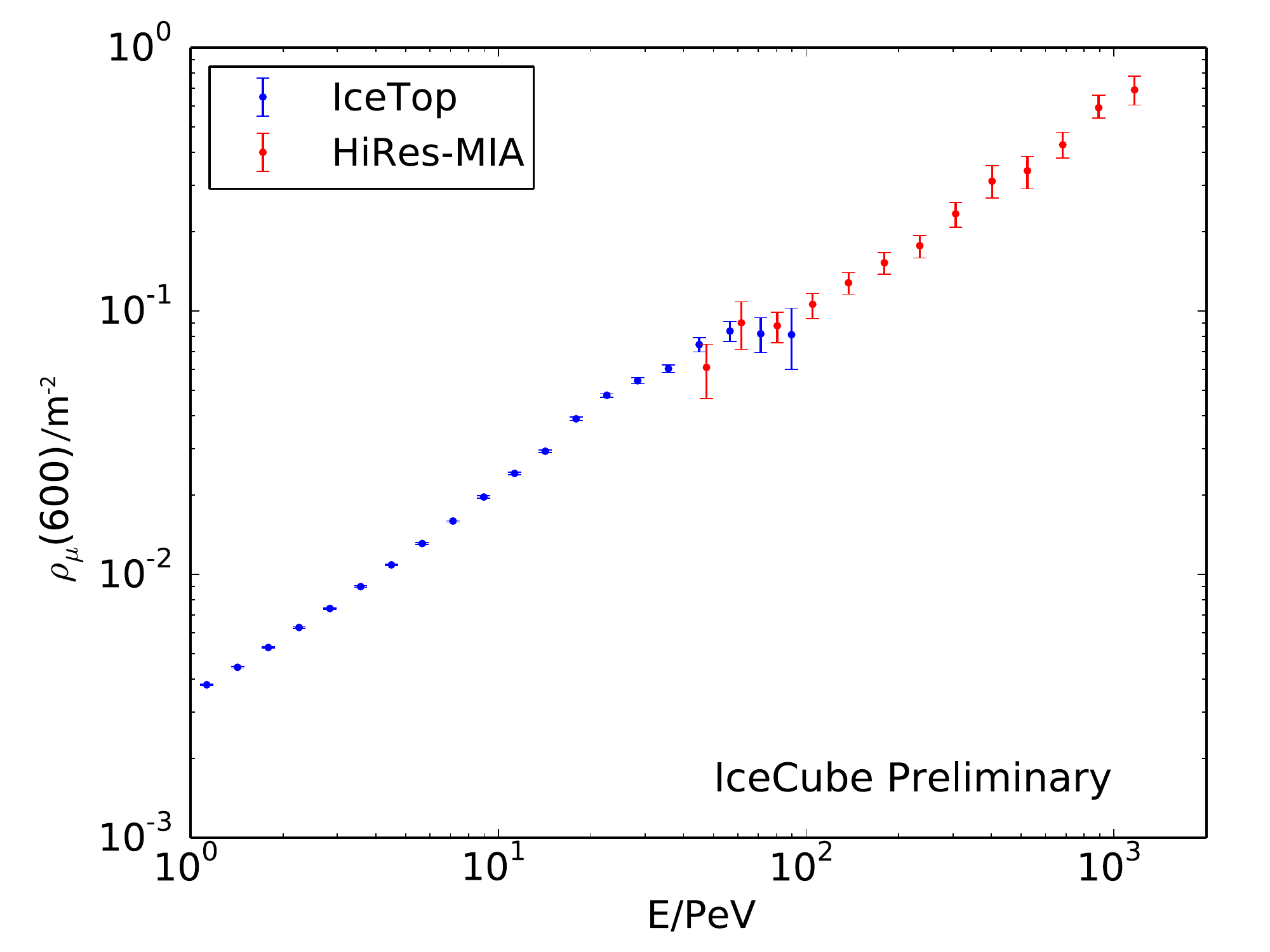}
    \caption{The energy dependence of the muon density
      $\rho_{\mu}(600)$ for vertical showers. Also shown is the result
      from the HiRes-MIA collaboration \cite{PhysRevLett.84.4276}.}
    \label{fig:hires_mia_comparison}
  \end{center}
\end{figure}

The resulting muon density at \SI{600}{m} from the shower axis is
displayed in Figure~\ref{fig:hires_mia_comparison} for the case of
vertical events. Here one can see that the muon density is in the same
order of magnitude as the Hires-MIA experiment
\cite{PhysRevLett.84.4276}. One remarkable result is that the
dependence on energy agrees very well with the Akeno air shower array
result \cite{Akeno_1984, Akeno_muons_1995}, which reported a power law
dependence with an index of 0.83.

The interpretation of any apparent differences in the absolute scale in
Figure~\ref{fig:hires_mia_comparison} requires some care.  We must
remember that the MIA array was not located at the same atmospheric
depth as IceTop (the depths are 860 and \SI{680}{g/cm^2}
respectively), which means that the densities need to be corrected to
account for attenuation in the atmosphere. Additionally, the detection
thresholds are slightly different. However, it is very encouraging
that the apparent offset in Figure~\ref{fig:hires_mia_comparison}
is on the order of the expected attenuation.

At this moment we do not believe that any deviation from a power law
behavior in Figure~\ref{fig:hires_mia_comparison} should be given much
importance. The cause for such deviations can still be due to
limitations of our method or the choice of lateral distance cut. These
and other systematic effects are currently under study.

A more detailed comparison with the expectation from simulations will
remain for future contributions, after we have taken into
consideration all relevant systematic effects. In this contribution we
note that the statistical errors in the present measurement
are significantly smaller than any other previous measurement. We
expect to make a statement on the absolute scale of the
muon density at large lateral distances in the near future.

We are considering improvements to the current analysis. The main
limitation at energies larger than $\sim$~\SI{30}{PeV} arises from our
selection of signals beyond a given lateral distance cut, together with the finite
size of the array. These two constraints limit the lever arm in the lateral
fit. One of the possibilities we are considering is the extension of
this analysis to include air showers that go through the detector deep
in the ice but not through IceTop, which would increase our lateral
distance range. We could also derive improvements from changes in our
lateral distance cut.

An obvious application of this measurement will be the improvement
of our event-by-event reconstruction to include a muon LDF together
with an EM LDF, which we expect will increase our sensitivity to the
mass of the primary. This method will provide a measure of primary
composition that is independent of the current approach of measuring
the high-energy muon bundles deep in the ice ($E_{\mu} \gtrapprox 300$
GeV). We expect that a comparison of these two independent measures
can help in identifying systematic effects arising from different
hadronic models used in Monte Carlo simulations of air showers.

\end{document}